\documentclass[referee]{raa}            

\UseRawInputEncoding
\usepackage{graphicx}             
\usepackage{times}
\usepackage{natbib}
\usepackage{appendix}
\usepackage{amsmath}
\usepackage{amssymb}
\usepackage{epsfig}
\usepackage{rotating}
\usepackage{color}
\usepackage{longtable}
\usepackage{pdflscape,lscape}
\usepackage{float}
\usepackage{placeins}
\usepackage{booktabs}
\usepackage{upgreek}
\usepackage{url}
\usepackage{hyperref}
\hypersetup{colorlinks = true, linkcolor = green, anchorcolor = red, citecolor = blue, filecolor = red, pagecolor = red, urlcolor = red}
\begin{document}

   \title{Analysis of TESS field eclipsing binary star V948\,Her: a pulsating or non-pulsating star?
}

   \volnopage{Vol.0 (20xx) No.0, 000--000}      
   \setcounter{page}{1}          

   \author{F. Kahraman Ali\c{c}avu\c{s}
      \inst{1,2}
   \and \"{O}. Ekinci
      \inst{2,3}
   }

   \institute{\c{C}anakkale Onsekiz Mart University, 
Faculty of Sciences and Arts, Physics Department, 17100, \c{C}anakkale, Turkey; {\it filizkahraman01@gmail.com}\\
        \and
             \c{C}anakkale Onsekiz Mart University, Astrophysics Research Center and Ulup\u0131nar Observatory, TR-17100, Çanakkale, Turkey\\
             \and
             \c{C}anakkale Onsekiz Mart University, School of Graduate Studies, Department of Space Sciences and Technologies, TR-17100, Çanakkale, Turkey\\
\vs\no
   {\small Received~~20xx month day; accepted~~20xx~~month day}}

\abstract{Pulsating stars occupy a significant place in the H-R diagram and it was thought that all stars inside the classical instability strip should pulsate. However, recent studies showed that there are many non-pulsating stars placed inside the classical instability strip. The existence of these non-pulsating stars is still a mystery. To deeply understand the properties of these non-pulsating and pulsating stars, one needs precise fundamental stellar parameters (e.g mass). For this purpose, the eclipsing binaries are unique systems. Hence, in this study, we present the TESS data analysis of one candidate pulsating eclipsing binary system V948\,Her. TESS data were used for the binary modelling with the literature radial velocity measurements and the precise fundamental parameters of the system were obtained. The system's age was derived as 1$\pm$0.24 Gyr. The positions of the binary components in the H-R diagram were examined and the primary component was found inside the $\delta$\,Scuti instability strip. However, in the frequency analysis of TESS data, we found no significant pulsation frequencies. Only the harmonics of the orbital periods were obtained in the analysis. Therefore, the system was classified as a non-pulsator. V948\,Her is an important object to understand the nature of non-pulsating stars inside the $\delta$\,Scuti instability strip.
 }

   \authorrunning{Kahraman Ali\c{c}avu\c{s} \& Ekinci}            
   \titlerunning{V948\,Her: a pulsating or non-pulsating star?}  

   \maketitle
   
\section{Introduction}           
\label{sect:intro}

Space-based observations have significantly contributed to astrophysical investigations. In particular, recent space missions such as CoRoT (\citeauthor{2006cosp...36.3749B}, \citeyear{2006cosp...36.3749B}), \textit{Kepler} (\citeauthor{2010Sci...327..977B}, \citeyear{2010Sci...327..977B}) and Transiting Exoplanet Survey Satellite (TESS, \citeauthor{2014SPIE.9143E..20R} \citeyear{2014SPIE.9143E..20R})  have provided high-quality photometric data of many stellar objects. Thanks to these data, the characteristics of many stellar systems have been revealed, while some new mysteries about them have come along. 

High-quality space data have created a revolution in the studies of pulsating and binary stars. The binary systems, particularly the eclipsing binary stars, are unique systems to directly and sensitively derive the fundamental stellar parameters of mass ($M)$ and radius ($R$). These fundamental stellar parameters could be obtained with an accuracy of $\sim$1\% with the simultaneous analysis of light and radial velocity ($v_{r}$) curves (\citeauthor{2010A&ARv..18...67T}, \citeyear{2010A&ARv..18...67T}). The spaced-based photometric observations have increased the sensitivity on these parameters with their high-precision data (\citeauthor{2020MNRAS.498..332M}, \citeyear{2020MNRAS.498..332M}). The precise fundamental stellar parameters are substantial for an extensive examination, understanding of the stellar evolution and calibrating the theoretical models (\citeauthor{2020A&A...637A..60T}, \citeyear{2020A&A...637A..60T}). In eclipsing binary systems, these space-based data have also disclosed some out-of-eclipse variations which could not be detected from ground-based studies. Those variations could be the results of stellar spots and/or pulsations in the binary component(s) (\citeauthor{2020ApJ...905...67P}, \citeyear{2020ApJ...905...67P}; \citeauthor{2020MNRAS.498..332M}, \citeyear{2020MNRAS.498..332M}). Additionally, it was also shown that these out-of-eclipse variations could be the effect of the Doppler-beaming caused by the orbital motion of the binary components (\citeauthor{2014A&A...563A.104H}, \citeyear{2014A&A...563A.104H}; \citeauthor{2018MNRAS.480.3864E}, \citeyear{2018MNRAS.480.3864E}). A comprehensive investigation of these out-of-eclipse variations could give us more information about the stellar systems such as about their interior structure with the help of pulsation modes. 

Pulsating binary systems have been known for decades. These systems occupy a notable place in astrophysical investigations because of their pulsation and eclipsing characteristics. The research of pulsations provides us information about the interior structure of stars, while studies of eclipsing binaries help us to determine the precise fundamental parameters. Therefore, the combination of these two properties creates a particular tool in investigating stellar structure and evolution. Examinations of high-quality space-based data have shown us that the pulsation characteristics of some oscillation stars could not have been fully discerned yet (\citeauthor{2014ApJ...796..118A}, \citeyear{2014ApJ...796..118A}; \citeauthor{2018MNRAS.476.3169B}, \citeyear{2018MNRAS.476.3169B}). There are some open questions especially about the A-F type $\delta$\,Scuti and $\gamma$\,Doradus pulsators. In the study of \citeauthor{2011A&A...534A.125U} (\citeyear{2011A&A...534A.125U}), about 750 \textit{Kepler} field $\delta$\,Scuti, $\gamma$\,Doradus stars and their hybrids were studied and it turned out that there are some pulsating stars located beyond their own instability strips. This result conflicts with the theoretical studies because theoretically, these pulsating stars should not exhibit their kind of oscillations out of their own domain. The other interesting thing is that some non-pulsators were found inside the instability strip of these oscillating stars (\citeauthor{2011A&A...534A.125U} \citeyear{2011A&A...534A.125U}). M. Breger suggested that probably all stars inside the classical instability strip show oscillations but some of them have a very low amplitude which could not be detected from the ground-based observations because of the accuracy of the observations (\citeauthor{1969ApJS...19...79B} \citeyear{1969ApJS...19...79B}). However, recent studies with space-based photometric data showed that there are non-pulsators inside the instability strips of these variables (\citeauthor{2013AstRv...8c..83G}, \citeyear{2013AstRv...8c..83G}; \citeauthor{2015MNRAS.447.3948M}, \citeyear{2015MNRAS.447.3948M}; \citeauthor{2019MNRAS.489.3285N}, \citeyear{2019MNRAS.489.3285N}). In a recent study, \citeauthor{2019MNRAS.485.2380M}, (\citeyear{2019MNRAS.485.2380M}), examined the fraction of pulsators inside the $\delta$\,Scuti instability strip. This fraction amount was found to be decreasing through the red and blue edges, while it reaches 70\% in the middle of the $\delta$\,Scuti instability strip. It was clearly shown that there are some stars in the $\delta$\,Scuti instability strip and they exhibit no oscillations at the micro-magnitude sensitivity level of TESS data. The secret of these non-pulsators still remains, although the studies were carried out about these systems (\citeauthor{2011MNRAS.417..591B}, \citeyear{2011MNRAS.417..591B}; \citeauthor{2013AstRv...8c..83G}, \citeyear{2013AstRv...8c..83G}; \citeauthor{2015MNRAS.447.3948M}, \citeyear{2015MNRAS.447.3948M}; \citeauthor{2019MNRAS.489.3285N}, \citeyear{2019MNRAS.489.3285N}). To understand the mysteries of these non-pulsators, the eclipsing binary systems with non-pulsating or pulsating component(s) inside the  
$\delta$\,Scuti instability strip could be useful. Therefore, in this study, we focused on the analysis of an eclipsing binary system, V948\,Her, which is believed to have a candidate $\delta$\,Scuti binary component. 

The eclipsing variability of V948\,Her was found by \textit{Hipparcos} (\citeauthor{1997ESASP1200.....E}, \citeyear{1997ESASP1200.....E}). The primary component of the system was listed as a candidate pulsator by \citeauthor{2006MNRAS.370.2013S} (\citeyear{2006MNRAS.370.2013S}). \citeauthor{2012NewA...17..634L} (\citeyear{2012NewA...17..634L}) classified the system as a detached eclipsing binary and search for possible pulsations by using ground-based observations. However, they found no convincing result. A comprehensive study of V948\,Her was given by \citeauthor{2018RAA....18...87K} (\citeyear{2018RAA....18...87K}). In this study, the ELODIE spectra and the Super Wide Angle Search for Planets (SuperWASP) photometric data were used. The system was found to be a single-lined binary and the atmospheric parameters, projected rotational velocity ($v \sin i$) and the metallicity (Fe/H) of the primary component were obtained by using the available spectra. Simultaneous analysis of $v_{r}$ and light curve was presented. The author also pointed out that some oscillation-like behaviours in the SuperWASP data are available and more observations are needed to confirm this variability. To reveal and understand the out-of-eclipse variations in this system, high-precision photometric data has been needed. Thanks to the TESS mission now the system has high-quality photometric data and this allows us to probe the out-of-eclipse variations in this system, obtain precise fundamental stellar parameters and examine the evolution of the system. Therefore, a revised analysis of V948\,Her is presented in this study. The paper is organised as follows. The information about the photometric data and new ephemeris calculation are given in Sect.\ref{sect:Obs}. The binary modelling and the time-series analysis are presented in the Sect.\ref{sect:binarymodel}. The discussions and conclusion are presented in Sect.\ref{sect:conclusion}.

\section{Photometric data and new ephemeris calculation}
\label{sect:Obs}

V948\,Her was observed with the Transiting Exoplanet Survey Satellite (TESS). The TESS was originally designed 
for discovering new exoplanets orbiting around bright stars ($4<I{_c}<13$) in the sky. The TESS has carried out its missions by dividing the sky into partly overlapping 26 sectors. Each of these sectors was observed around 27-days with 2-mins short cadence (SC) and 30-mins long cadence (LC) observation modes in a wide bandpass ranging from 600 to 10000\,nm (\citeauthor{2014SPIE.9143E..20R} \citeyear{2014SPIE.9143E..20R}). The mission completed its 2-years primary missions and continues an extended mission now. 

\begin{figure}
   \centering
   \includegraphics[width=12cm, angle=0]{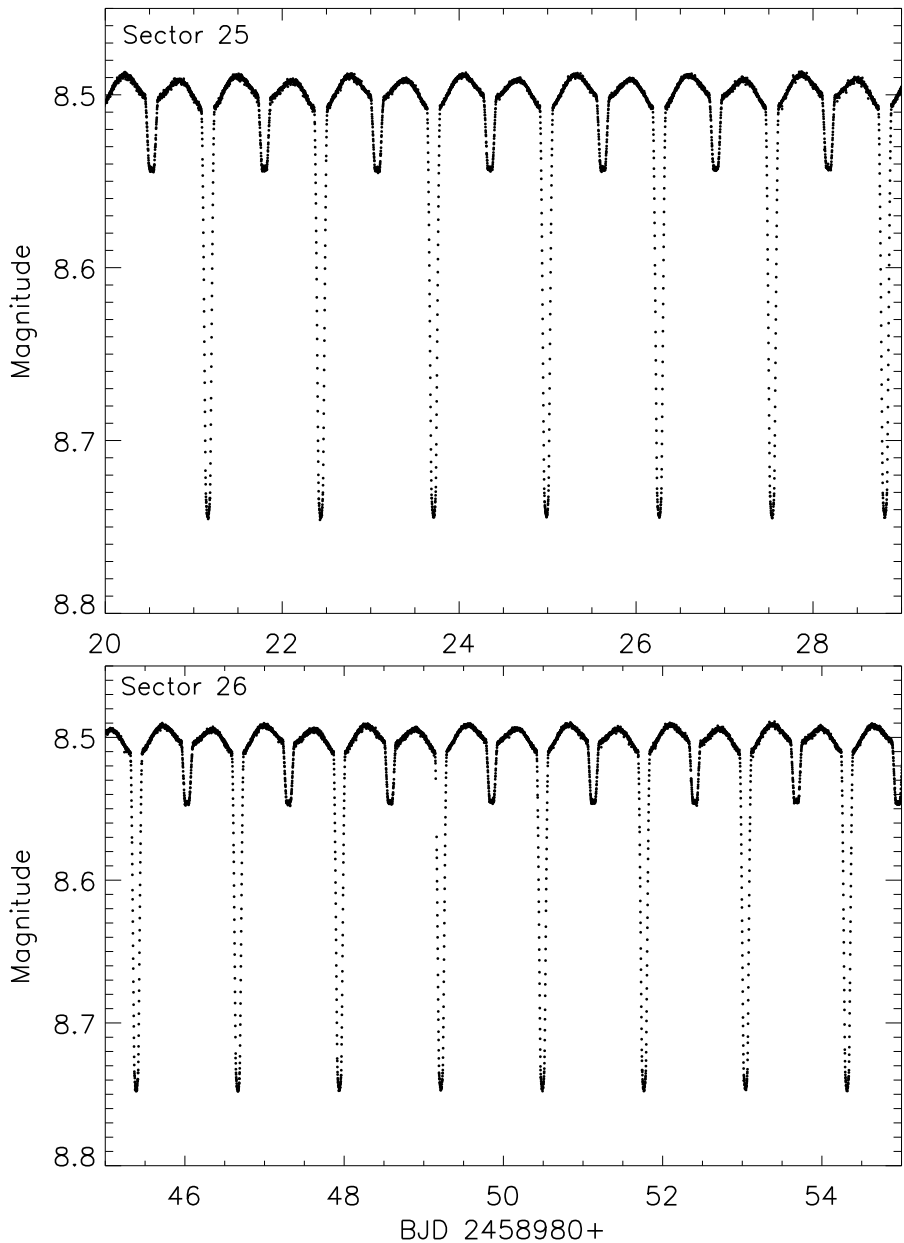}
   \caption{Light curve of $\sim$10-days from sector 25 and 26 observations.}
   \label{lc_original}
   \end{figure}

During the first two years, the TESS monitored V948\,Her in two continuous sectors 25 and 26 in 2-mins SC mode. In the study, we plan to seek for possible $\delta$\,Scuti type pulsations in the system, therefore all SC data were taken into account. The SC data are more suitable for the asteroseismology of $\delta$\,Scuti stars because its Nyquist frequency reaches around 340 d$^{-1}$. The SC data of V948\,Her were taken from the  Mikulski Archive for Space Telescopes (MAST) archive \footnote{https://mast.stsci.edu/}. The PDC$\_$SAP fluxes were converted into magnitudes and the scattered points over 2$\sigma$
were cleaned from the data. The PDC$\_$SAP data of the sector 25 and 26 for around 10 continuous days are shown in Fig.\ref{lc_original}. As clearly seen from the figure, there are asymmetries on the out-of-eclipse light curve and these could be the effect of a spot(s) placing on the binary component(s).

The system was also observed at the \c{C}anakkale Onsekiz Mart
University Observatory with the Apogee ALTA U47 CCD camera placed on the 30-cm telescope to obtain a new minima time. We observed the primary minimum in 9 July 2021 with V-band. The photometric data were reduced by following the classical reduction steps (bias, dark subtraction and flat correction) with the C-Munipack\footnote{http://c-munipack.sourceforge.net/} software. The new HJD minima time was calculated to be 2459405.40676\,$\pm$\,0.0002 using the method of \citeauthor{1956BAN....12..327K} (\citeyear{1956BAN....12..327K}). Additionally, we calculated the minima times from the TESS and also available SuperWASP data of the system to examine its period variations. The primary and secondary minima times were derived from the first, middile and last part of the each TESS sectors. In total twelve TESS minima times were obtained. We also gathered three primary minima times from the SuperWASP data. The literature minima times of V948\,Her were also collected from the O-C Gateway\footnote{http://var2.astro.cz/ocgate/}. In total, we used 26 minima times and investigated the period changes of the system. All used minima times are listed in Table\,\ref{minima_table}. Those minima times were first investigated with the method of \citeauthor{2009NewA...14..121Z} (\citeyear{2009NewA...14..121Z}).
 According to our initial examination, we found that the system shows a 1.13-second period decrease in one hundred years. However, because of the fewer minima times, this result was not reliable and the system needs more minima times to obtain more decent results. For this reason, in this study, we only determined new light elements of V948\,Her by fitting a linear fit to all minima times. The new light elements are given in the following equation:

\begin{equation}
\centering
 HJD (MIN I)= 2459405.4012(7) + 1.2752049E(3)
\end{equation}

\section{Binary modelling and Frequency analysis}
\label{sect:binarymodel}
In the binary modelling of V948\,Her, we controlled the level of maximum points in all available TESS SC data to check whether they show differences from one sector to the other. No significant variation was found. Therefore, we used only the data of sector 25 in the binary modelling. These data were binned to 4000 points. The Wilson-Devinney code (\citeauthor{1971ApJ...166..605W}, \citeyear{1971ApJ...166..605W}) with the Monte-Carlo simulation (\citeauthor{2004AcA....54..299Z}, \citeyear{2004AcA....54..299Z}; \citeauthor{2010MNRAS.408..464Z}, \citeyear{2010MNRAS.408..464Z}) was used in the study to derive reliable uncertainties in the searched parameters.

In the binary modelling, the effective temperature ($T_{\rm eff}$) of the primary (hot) binary component was taken to be 7100\,$\pm$\,200\,K from the results of the literature spectral analysis (\citeauthor{2018RAA....18...87K}, \citeyear{2018RAA....18...87K}) and this value was fixed during the analysis. The other fixed parameters are the bolometric albedos, bolometric gravity-darkening coefficient and the logarithmic limb darkening coefficient and they were taken the same as given in the study of \citeauthor{2018RAA....18...87K} (\citeyear{2018RAA....18...87K}). The previous binary modelling of the system was done with the SuperWASP data. As the TESS data is quite sensitive comparing the SuperWASP, the binary parameters could be derived more accurately. Therefore, in the binary modelling, we searched for important binary parameters such as orbital inclination ($i$), the mass ratio ($q$), surface potential ($\Omega$) of the binary component and the $T_{\rm eff}$ of the cool secondary component. As V948\,Her is defined as a detached binary system (\citeauthor{2012NewA...17..634L}, \citeyear{2012NewA...17..634L}; \citeauthor{2018RAA....18...87K}, \citeyear{2018RAA....18...87K}), a detached binary configuration was used in the modelling. 

\begin{figure}
   \centering
   \includegraphics[width=12cm, angle=0]{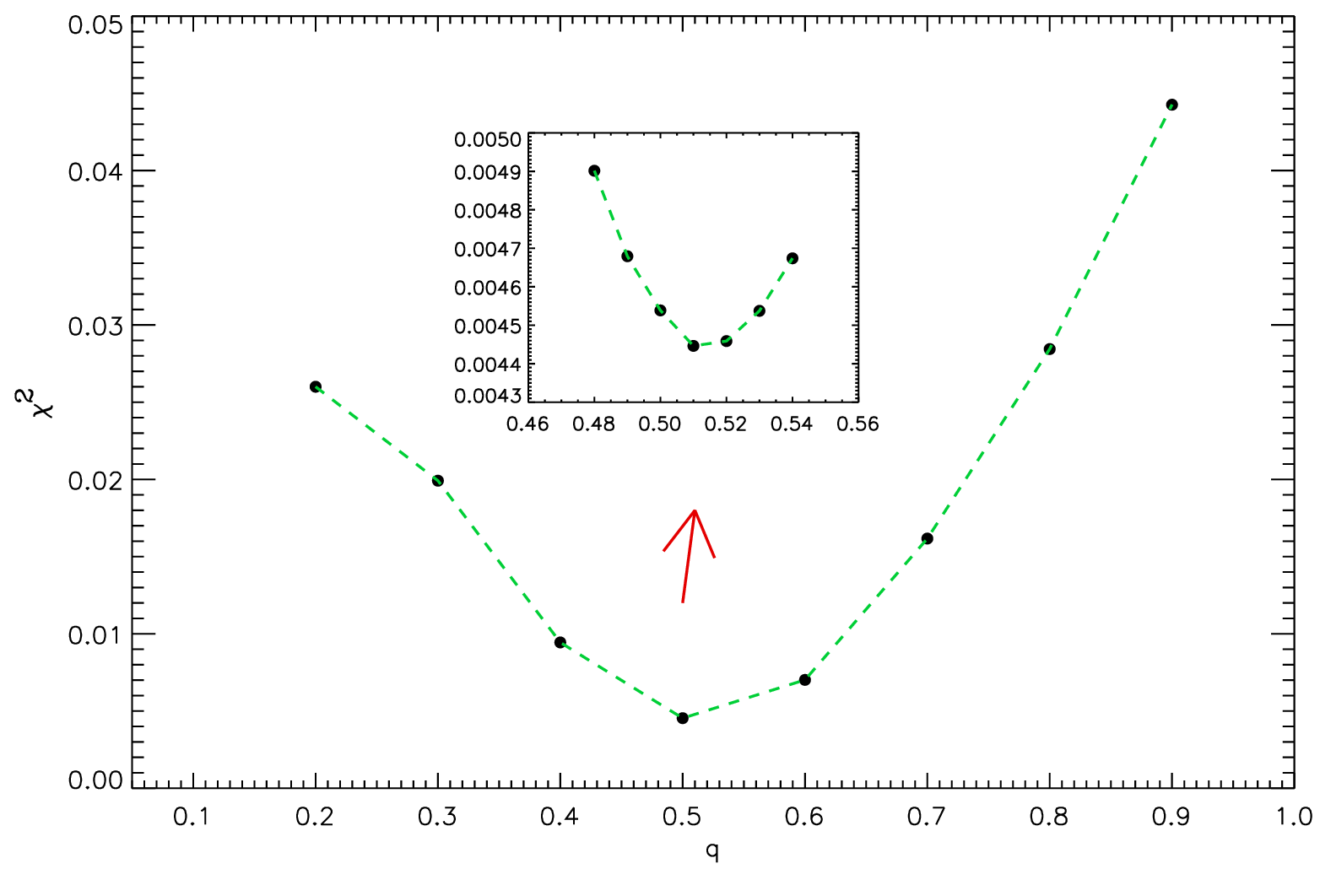}
   \caption{Results for the $q$ search.}
   \label{q_search}
   \end{figure}

 \begin{figure}
   \centering
   \includegraphics[width=12cm, angle=0]{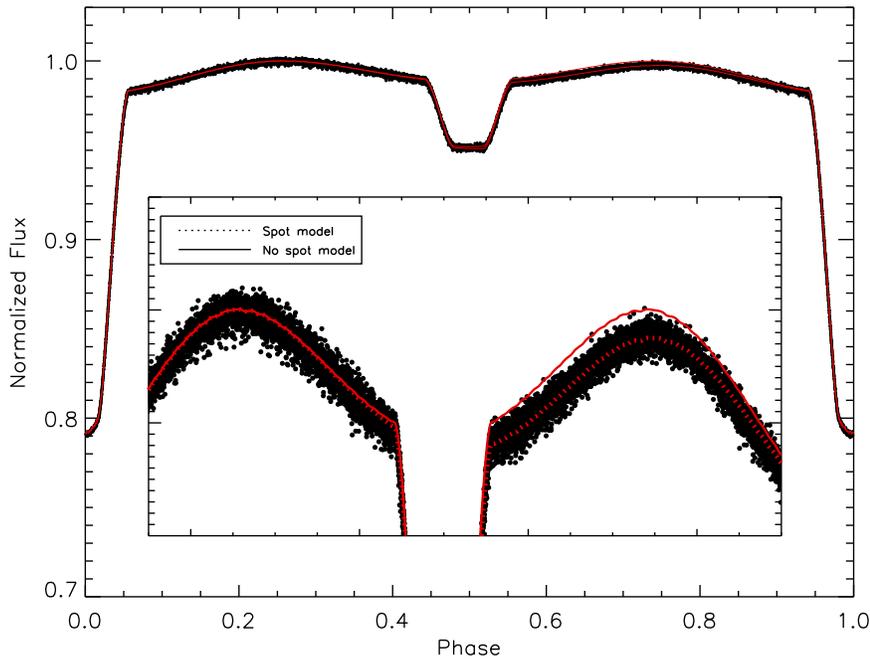}
   \caption{Theoretical fits (red lines) to the TESS phased data and the consistency of the spotted model (dotted line).}
   \label{lcfit}
   \end{figure}
   
We first carried out a model without spot assumption on the system. However, the theoretical models without spot did not represent the binary variation. Therefore, a spot was included in the binary modelling. First, a cool spot was assumed on the surface of the secondary cool star however no significant results were obtained. Hence, new binary modelling was carried out by assuming a cool spot on the primary component. We obtained a better $\chi^2$ value for the modelling with a cool spot on the surface of the primary component. As the system is a single-lined binary, we do not have precise information about the $q$ value. Therefore, we carried out a $q$ search to find the $q$ value corresponding to the minimum $\chi^2$. The result of this search is shown in Fig.\ref{q_search}. We obtained the minimum $\chi^2$ for the 0.51. Taking this $q$ value as input, binary modelling was performed again and the results of this modelling are given in Table\,\ref{lcresult}. The theoretical binary models with and without the spot are illustrated in Fig.\,\ref{lcfit}.
   
\begin{table}
\begin{center}
\caption[]{Results of the Light Curve Analysis and the Fundamental Stellar Parameters. The Subscripts 1, 2 and 3 Represent the Primary, the Secondary, and Third Binary Components, Respectively. $^a$ Shows the Fixed Parameters.}\label{lcresult}
 \begin{tabular}{lrr}
  \hline\noalign{\smallskip}
   Parameter			  &  Value   	   &  Value    \\
                          & (This Study)     & (\citeauthor{2018RAA....18...87K}, \citeyear{2018RAA....18...87K}) \\
  \hline\noalign{\smallskip}
$i$ ($^{o}$)	    & 85.518 $\pm$ 0.031 &  86.86 $\pm$ 0.044    \\	
$T$$_{1}$$^a$ (K)   & 7100 $\pm$ 200  & 7100 $\pm$ 200	   	\\	
$T$$_{2}$ (K)       & 4280 $\pm$ 230  & 4430 $\pm$ 250 \\
$\Omega$$_{1}$		& 4.516 $\pm$ 0.004 & 4.521 $\pm$ 0.005 \\
$\Omega$$_{2}$		& 5.942 $\pm$ 0.006 & 5.584 $\pm$ 0.010 \\
Phase shift         & -0.0011 $\pm$ 0.0001 &   0.00040 $\pm$ 0.00004	  \\
$q$                 & 0.504 $\pm$ 0.001  & 0.443 $\pm$ 0.001 \\
$r$$_{1}$$^*$ (mean) & 0.2512 $\pm$ 0.0036 &   -  \\
$r$$_{2}$$^*$ (mean) & 0.1075 $\pm$ 0.0026 &   -   \\
$l$$_{1}$ / ($l$$_{1}$+$l$$_{2}$)  & 0.970 $\pm$ 0.013 & 0.981 $\pm$ 0.004\\  
$l$$_{2}$ / ($l$$_{1}$+$l$$_{2}$)  & 0.030 $\pm$ 0.016 & 0.019 $\pm$ 0.005\\  
$l$$_{3}$             & 0.00   &  0.00\\  
\multicolumn{2}{c}{Spot Parameters}\\

Co-Latitude (deg)  & 90$^a$ & -  \\
Longitude (deg)   & 114.340$\pm$ 0.112 & -\\
Radius  (deg)          & 25.720$\pm$0.038 & -\\
Temperature Factor$^{**}$  & 0.995$\pm$0.003 & -\\
\multicolumn{2}{c}{Derived Quantities}\\

$M$$_{1}$ ($M_\odot$)	 & 1.274 $\pm$ 0.012 & 1.722 $\pm$ 0.123   \\	
$M$$_{2}$ ($M_\odot$)	 & 0.642 $\pm$ 0.014 & 0.762 $\pm$ 0.020   	\\
$R$$_{1}$ ($R_\odot$)	 & 1.543 $\pm$ 0.022 & 1.655 $\pm$ 0.034 	  \\
$R$$_{2}$ ($R_\odot$)	 & 0.661 $\pm$ 0.016 & 0.689 $\pm$0.016  \\
log ($L$$_{1}$/$L_\odot$) & 0.737 $\pm$ 0.028 & 0.797 $\pm$ 0.034  \\
log ($L$$_{2}$/$L_\odot$) & -0.880 $\pm$ 0.034 & -0.783 $\pm$ 0.016  \\
$\log g$$_{1}$ (cgs)      & 4.16 $\pm$ 0.02 & 4.23 $\pm$ 0.10   \\
$\log g$$_{2}$ (cgs)      & 4.60 $\pm$ 0.03 & 4.63 $\pm$ 0.10   \\
$M_{bol}$$_{1}$ (mag)     & 2.898 $\pm$ 0.082 & 2.763 $\pm$ 0.691	  \\
$M_{bol}$$_{2}$ (mag)	  & 6.939 $\pm$ 0.104 & 6.715 $\pm$1.327      		  \\
$M_{V}$$_{1}$ (mag)	      & 2.820 $\pm$ 0.086  &   - 		  \\
$M_{V}$$_{2}$ (mag)	      & 7.534 $\pm$ 0.113  &    - 		  \\
\noalign{\smallskip}\hline
\end{tabular}
\end{center}
 \begin{description}
     \centering
 \item[ ]* fractional radii, ** $T_{\rm eff}$ $_{spot}$ / $T_{\rm eff}$ $_{star}$
 \centering
 \end{description}
\end{table}

The update binary modelling of V948\,Her show differences from the literature binary modelling (\citeauthor{2018RAA....18...87K}, \citeyear{2018RAA....18...87K}). The results of the previous binary modelling is also given in Table\,\ref{lcresult} for a comaprison. Especially the $\Omega$, $q$ and fractional radius ($r$) differ significantly. These affect the resulting fundamental parameters. The discrepancy in the results of this and the previous binary modelling is expected because of the quality of used data in the analyses. That is why no spot was found in the previous binary modellings. 
   
   The fundamental stellar parameters $M$, $R$, luminosity ($L$), surface gravity ($\log g$) and bolometric ($M_{bol}$), absolute magnitudes ($M_{V}$) were calculated using the mass function found in the radial velocity analysis (\citeauthor{2018RAA....18...87K}, \citeyear{2018RAA....18...87K}) and the binarity parameters obtained in this study. The resulting fundamental stellar parameters are given in Table\,\ref{lcresult}. When the fundamental parameters are compared with the previously given ones in the literature (see Table\,\ref{lcresult}), one could see that there are significant differences in those parameters and that would change the evolutionary scenario of the system and its position in the H-R diagram.  

   
   \begin{figure}
 \begin{subfigure}{0.9\textwidth}
  \centering
  \includegraphics[width=1.0\linewidth]{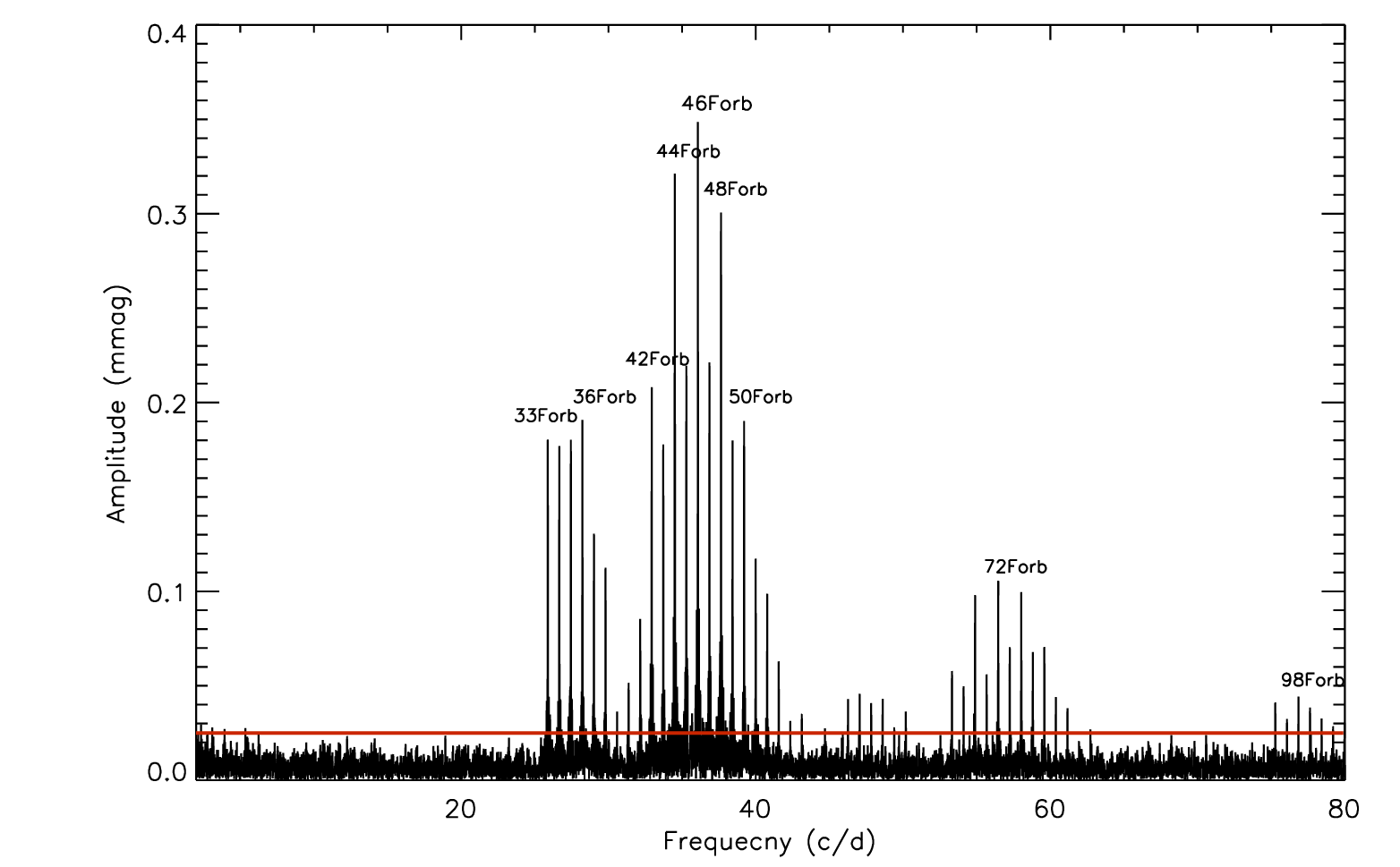}  
  \label{fig:sub-first}
\end{subfigure}
\begin{subfigure}{0.9\textwidth}
  \centering
  \includegraphics[width=1\linewidth]{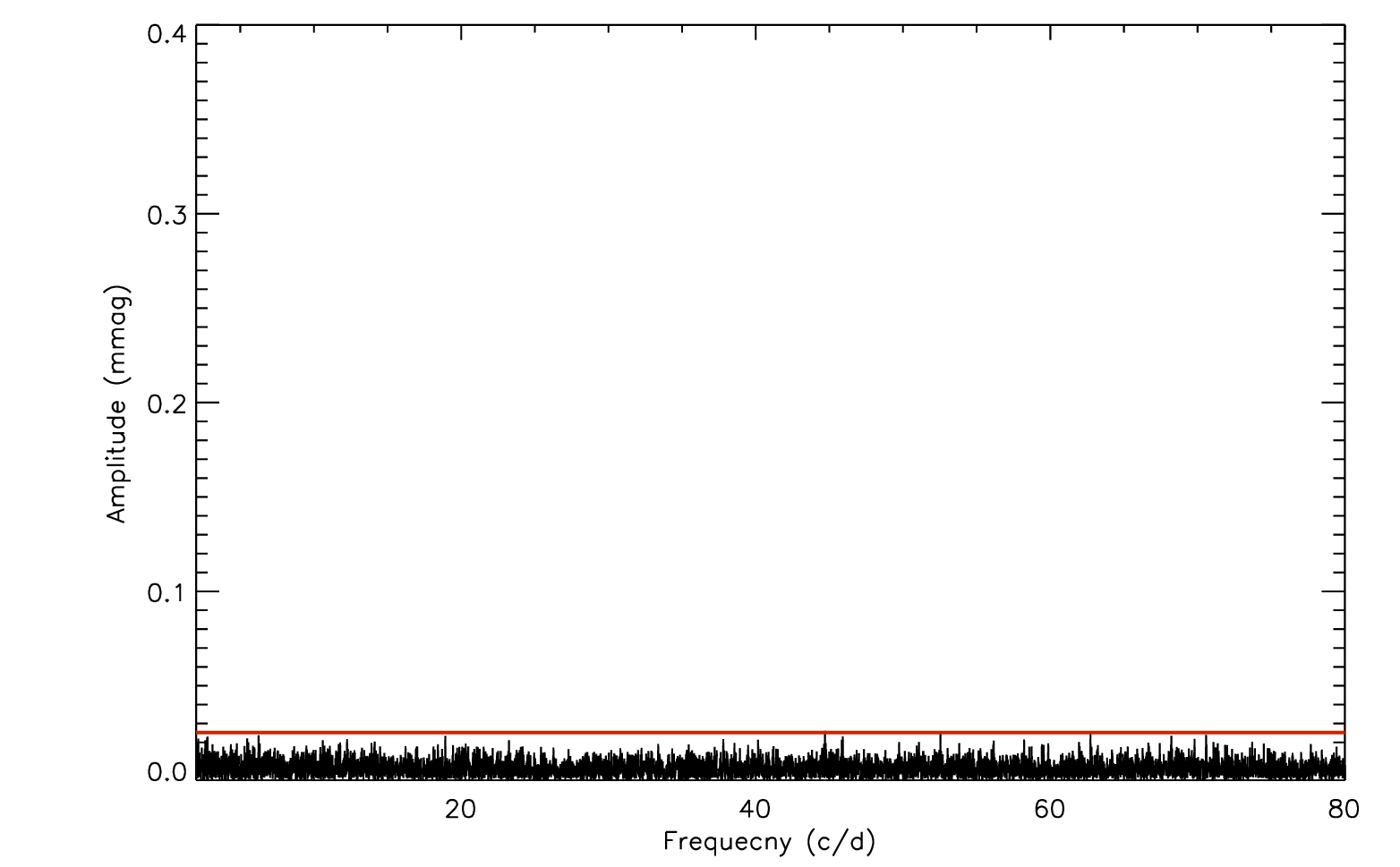}  
  \label{fig:sub-second}
\end{subfigure}
\caption{The Fourier spectrum of V948\,Her (top panel). The horizontal red line represents the 4-$\sigma$ level. The bottom panel represents the Fourier spectrum after prewhitening all significant harmonic frequencies of the orbital period.}
\label{fre}
\end{figure}

The primary component of V948\,Her was classified as a candidate $\delta$\,Scuti star (\citeauthor{2006MNRAS.370.2013S}, \citeyear{2006MNRAS.370.2013S}). In the study of \citeauthor{2012NewA...17..634L} (\citeyear{2012NewA...17..634L}), no convincing frequency with a signal-to-noise (S/N) ratio larger than 4 was found. However, in the study of \citeauthor{2018RAA....18...87K} (\citeyear{2018RAA....18...87K}), it was shown that the star is suspected and could show a pulsation frequency around 26 d$^{-1}$. In addition, it was pointed out that the system needs high-quality data to confirm the pulsational variability. Therefore, in this study, we present a frequency analysis to high-quality TESS data of V948\,Her. Before starting the analysis first the binary variation was removed from the light curve by fitting a 32-harmonics of the orbital period frequency of 0.784188 with the {\sc Period04} code (\citeauthor{2005CoAst.146...53L}, \citeyear{2005CoAst.146...53L}). The 32-harmonics of the orbital period frequency was chosen because it represents the binary variation well. The fit of the orbital period's harmonics was removed from the TESS data before searching for possible pulsation frequencies. A frequency analysis was performed to the residuals with the {\sc Period04} software that provides harmonics and combination frequencies. We carried out this analysis for the range of $0-80$ d$^{-1}$ until we reach the frequencies having S/N ratio smaller than 4 (\citeauthor{1993A&A...271..482B}, \citeyear{1993A&A...271..482B}). The resulting frequencies are listed in Table\,\ref{fre_table}. As can be seen from the table all frequencies are exactly the harmonics of the orbital period's frequency. The Fourier spectrum for this analysis is shown in Fig.\,\ref{fre}. Except for the harmonics of the orbital period no significant frequency with an S/N ratio over 4 was found as shown in the bottom panel of Fig.\,\ref{fre}. Therefore the system was defined as a non-pulsating star. With this result, we ruled out the pulsational variability of the primary component of V948\,Her.  

\begin{table}
\begin{center}
\caption[]{The resulting frequencies, amplitudes and the phases for the significant frequencies.}\label{fre_table}
 \begin{tabular}{lcccc}
  \hline\noalign{\smallskip}
       & Frequency    & Amplitude& Phase & S/N\\
        &  (d$^{-1}$) & (mmag)    & (rad) & \\
  \hline\noalign{\smallskip}
46$\nu_{orb}$ & 36.0728(2) & 0.358 & 0.573(2) &51.9 \\
44$\nu_{orb}$ & 34.5040(2) & 0.331 & 0.222(3) & 51.0\\
48$\nu_{orb}$ & 37.6416(2) & 0.308 & 0.924(3) & 43.1\\
47$\nu_{orb}$ & 36.8572(3) & 0.233 & 0.253(4) & 34.4\\
45$\nu_{orb}$ & 35.2884(3) & 0.232 & 0.908(4) & 33.1\\
42$\nu_{orb}$ & 32.9362(3) & 0.215 & 0.849(4) &31.9\\
36$\nu_{orb}$ & 28.2297(3) & 0.197 & 0.313(4) & 30.5\\
50$\nu_{orb}$ & 39.2094(3) & 0.196 & 0.299(4) &28.2\\
49$\nu_{orb}$ & 38.4250(3) & 0.188 & 0.630(5) & 28.6\\
43$\nu_{orb}$ & 33.7206(3) & 0.187 & 0.528(5) & 28.1\\
35$\nu_{orb}$ & 27.4463(3) & 0.186 & 0.601(5) & 33.2\\
33$\nu_{orb}$ & 25.8785(3) & 0.185 & 0.207(5) &32.5\\
34$\nu_{orb}$ & 26.6619(3) & 0.184 & 0.941(5) &32.0\\
37$\nu_{orb}$ & 29.0151(4) & 0.136 & 0.940(6) &20.1\\
51$\nu_{orb}$ & 39.9938(5) & 0.121 & 0.969(7) &17.5\\
38$\nu_{orb}$ & 29.7985(5) & 0.116 & 0.582(7) &17.4\\
72$\nu_{orb}$ & 56.4614(6) & 0.109 & 0.794(8) &15.9\\
74$\nu_{orb}$ & 58.0292(6) & 0.103 & 0.188(8) &15.5\\
52$\nu_{orb}$ & 40.7783(6) & 0.100 & 0.642(9) &14.2\\
70$\nu_{orb}$ & 54.8916(6) & 0.100 & 0.472(9) &14.0\\
41$\nu_{orb}$ & 32.1528(7) & 0.090 & 0.145(10) &12.4\\
73$\nu_{orb}$ & 57.2469(8) & 0.074 & 0.411(11) &11.2\\
76$\nu_{orb}$ & 59.5981(8) & 0.072 & 0.528(12) &12.3\\
75$\nu_{orb}$ & 58.8147(9) & 0.071 & 0.809(12) &10.9\\
53$\nu_{orb}$ & 41.5606(10) & 0.064 & 0.420(13) &9.3\\
71$\nu_{orb}$ & 55.6770(10) & 0.060 & 0.086(14) &8.6\\
68$\nu_{orb}$ & 53.3258(10) & 0.059 & 0.020(15) &8.5\\
40$\nu_{orb}$ & 31.3673(11) & 0.053 & 0.487(16) &7.5\\
69$\nu_{orb}$ & 54.1113(12) & 0.051 & 0.672(17) &7.4\\
60$\nu_{orb}$ & 47.0556(13) & 0.047 & 0.975(18) &7.1\\
98$\nu_{orb}$ & 76.8521(13) & 0.046 & 0.926(19) &7.7\\
77$\nu_{orb}$ & 60.3804(13) & 0.045 & 0.256(19) &7.5\\
62$\nu_{orb}$ & 48.6183(14) & 0.045 & 0.509(19) &7.2\\
59$\nu_{orb}$ & 46.2651(14) & 0.043 & 0.454(20) &5.9\\
96$\nu_{orb}$ & 75.2812(14) & 0.037 & 0.633(20) &5.5\\
61$\nu_{orb}$ & 47.8359(15) & 0.042 & 0.791(21) &6.3\\
99$\nu_{orb}$ & 77.6375(15) & 0.041 & 0.596(21) &6.4\\
78$\nu_{orb}$ & 61.1669(16) & 0.039 & 0.874(22) &6.0\\
39$\nu_{orb}$ & 30.5840(23) & 0.032 & 0.346(33) &5.0\\
64$\nu_{orb}$ & 50.1902(16) & 0.037 & 0.760(23) &5.1\\
97$\nu_{orb}$ & 76.0708(18) & 0.034 & 0.169(26) &5.6\\
100$\nu_{orb}$ & 78.4187(18) & 0.034 & 0.276(25) &5.4\\
55$\nu_{orb}$ & 43.1295(18) & 0.034 & 0.742(25) &5.1\\
102$\nu_{orb}$ & 79.9857(19) & 0.032 & 0.754(27) &4.6\\
54$\nu_{orb}$ & 42.3471(20) & 0.031 & 0.061(28) &4.8\\
101$\nu_{orb}$ & 79.2023(20) & 0.030 & 0.057(28) &4.8\\
63$\nu_{orb}$ & 49.4048(21) & 0.029 & 0.101(30) &4.5\\
\noalign{\smallskip}\hline
\end{tabular}
\end{center}  
\end{table}


\section{Discussions and Conclusion}
\label{sect:conclusion}

In this study, we present the TESS light curve analysis of V948\,Her. First, by using the high-quality TESS data a binary modelling was performed with the help of the literature radial velocity analysis's results (\citeauthor{2018RAA....18...87K}, \citeyear{2018RAA....18...87K}). Consequently, we found a best-fitting theoretical binary modelling with a spot on the surface of the primary component. By using the findings of the binary modelling and the results of the literature radial velocity analysis, the fundamental stellar parameters were calculated. These derived quantities are given in Table\,\ref{lcresult}. When these parameters especially the $M$ and $R$ values were compared with the literature ones, significant differences were seen. Especially the discrepancy in the $M$ values is around 4$\sigma$. That would affect the evolution of the system. 

\begin{figure}
   \centering
   \includegraphics[width=12cm, angle=0]{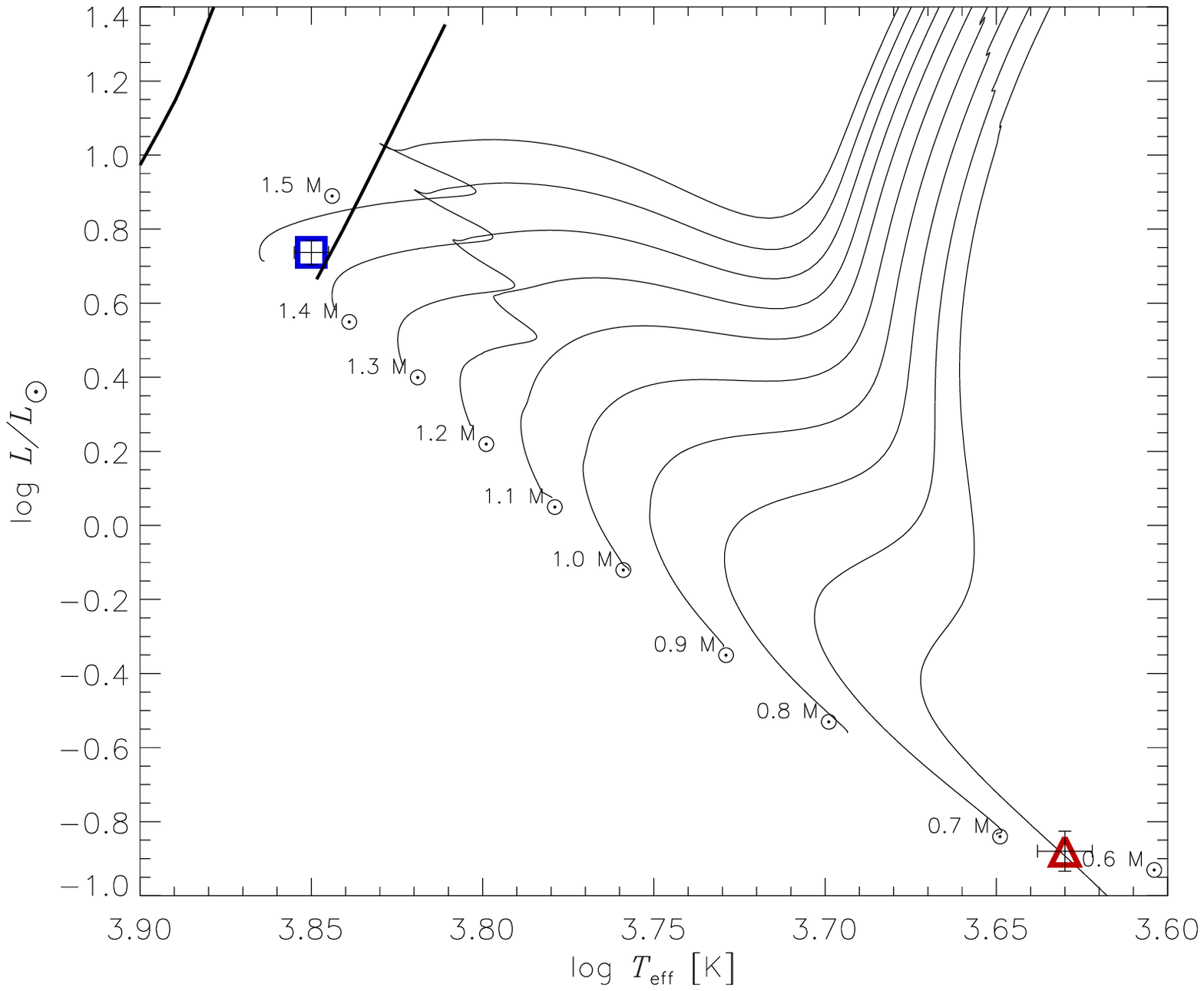}
   \caption{Position of the primary (square) and secondary (triangle) components of V948\,Her in the H-R diagram and the $\delta$\,Scuti instability strip (solid lines on the top-left) (\citeauthor{2005A&A...435..927D}, \citeyear{2005A&A...435..927D}).}
   \label{hr}
   \end{figure}

\begin{figure}
   \centering
   \includegraphics[width=12cm, angle=0]{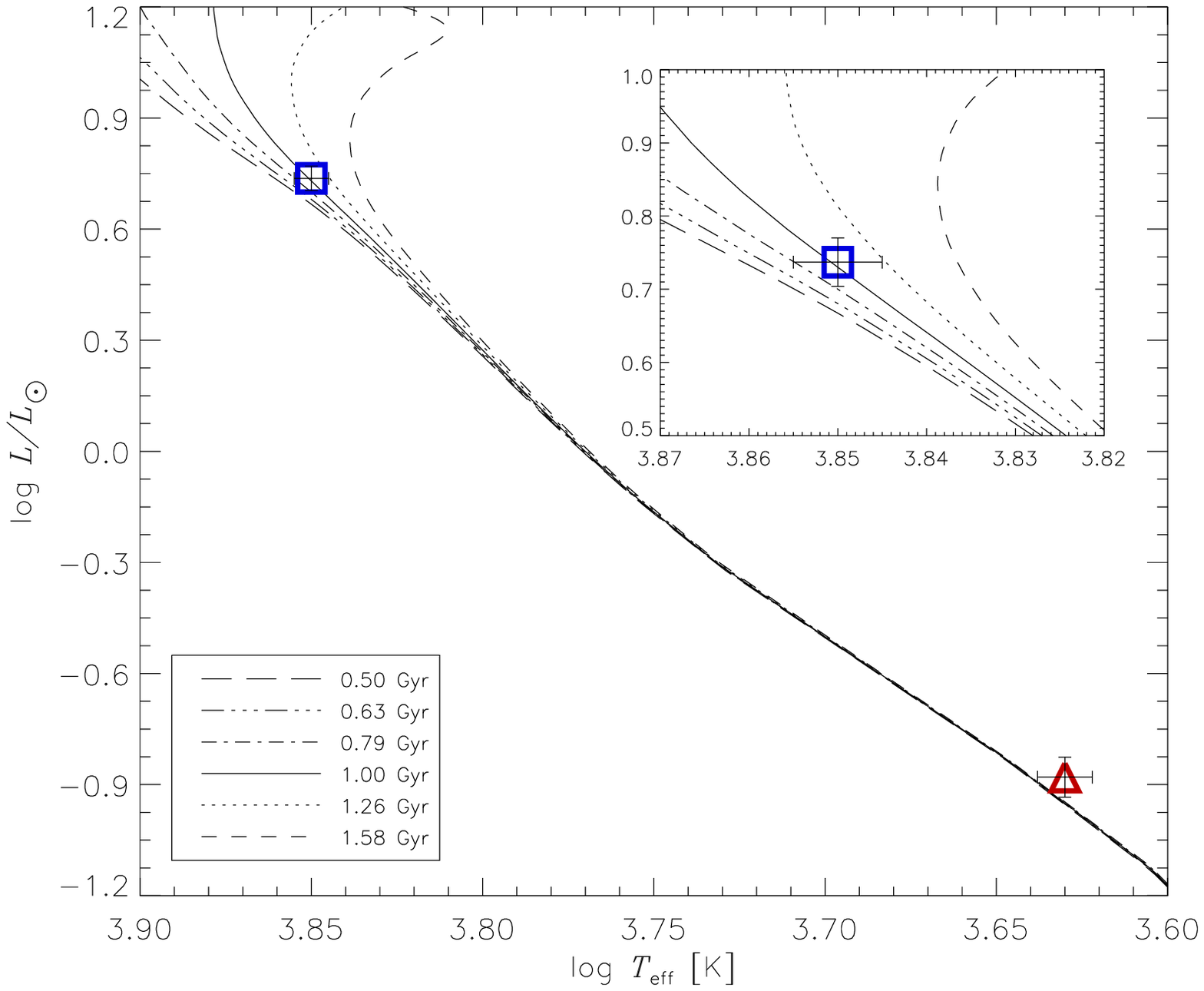}
   \caption{Isochrone for the primary (square) and secondary (triangle) components of V948\,Her.}
   \label{iso}
   \end{figure}

\begin{figure}
   \centering
   \includegraphics[width=12cm, angle=0]{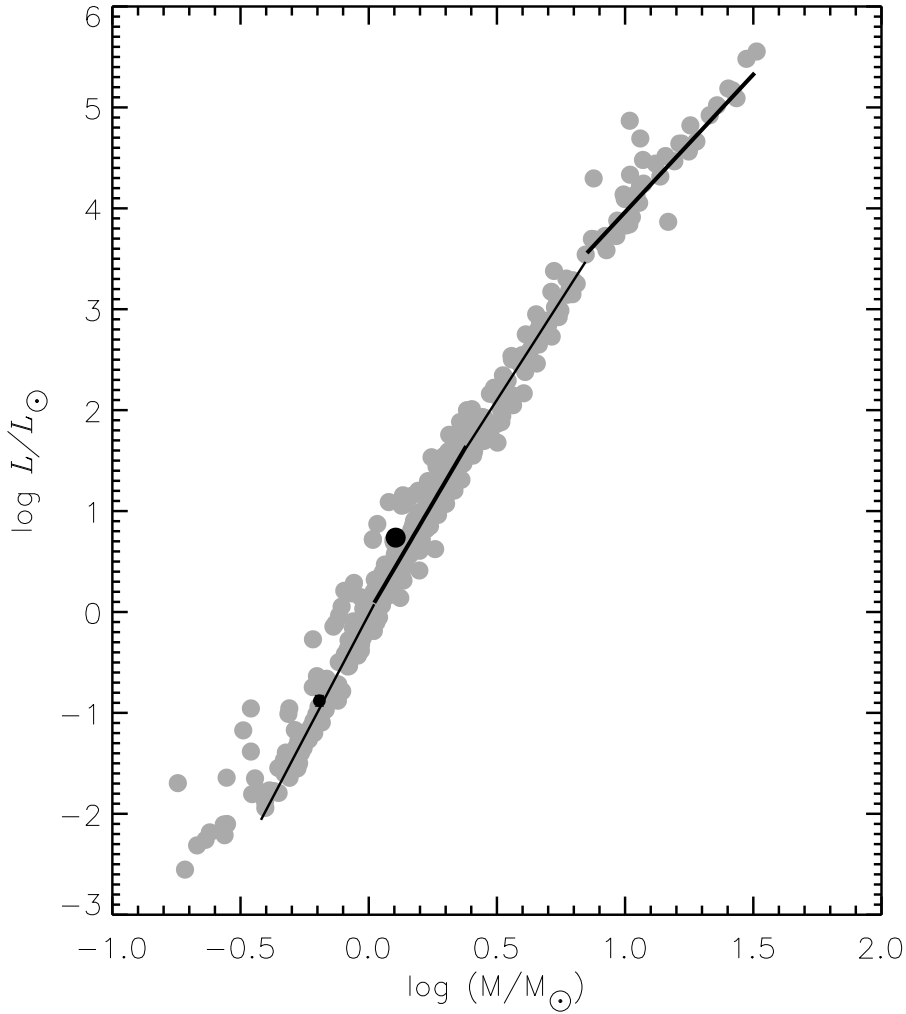}
   \caption{The position of the primary (bigger black dot) and secondary (smaller black dot) components of V948\,Her in the $\log$\,$M$\,$-$\,$\log$\,$L$ diagram. Gray circles represent the double-lined detached eclipsing binaries given by \citeauthor{2014PASA...31...24E} (\citeyear{2014PASA...31...24E}). 
The solid lines illustrate the relationships between the $L$ and $M$ (\citeauthor{2014PASA...31...24E}, \citeyear{2014PASA...31...24E}).}
   \label{ml}
   \end{figure}

By using the derived fundamental stellar parameters, the positions of both components in the Hertzsprung-Russell (H-R) diagram were examined. The evolutionary models were taken from the MESA Isochrones and Stellar Tracks (MIST\footnote{http://waps.cfa.harvard.edu/MIST/}) (\citeauthor{2016ApJS..222....8D}, \citeyear{2016ApJS..222....8D}; \citeauthor{2016ApJ...823..102C}, \citeyear{2016ApJ...823..102C}; \citeauthor{2015ApJS..220...15P}, \citeyear{2015ApJS..220...15P}; \citeauthor{2013ApJS..208....4P}, \citeyear{2013ApJS..208....4P}; \citeauthor{2011ApJS..192....3P}, \citeyear{2011ApJS..192....3P}). The evolutionary models were calculated for a mass range of 0.6$M_\odot$\,$<$\,$M$\,$<$\,1.5$M_\odot$. In the model calculations, a solar abundance (\citeauthor{2009ARA&A..47..481A}, \citeyear{2009ARA&A..47..481A}) was assumed considering the abundance analysis result of \citeauthor{2018RAA....18...87K} (\citeyear{2018RAA....18...87K}). The H-R diagram for the components of V948\,Her is illustrated in Fig.\ref{hr}. According to this diagram the primary component's $M$ value should be between 1.4 to 1.5\,$M_\odot$, however, it was found around 1.3\,$M_\odot$ in this study. This difference can not be found for the secondary component. It positions on the evolutionary model for 0.6\,$M_\odot$ which is in agreement with the secondary component's $M$ value found in this study. To estimate the system's age MESA isochrones were also used. For different ages, many isochrones with solar abundance were calculated. The positions of the primary and secondary components in the isochrone diagram are shown in Fig.\,\ref{iso}. Taking into account the primary component's parameters, we estimated the age of the system to be 1\,$\pm$\,0.24 Gyr. In Fig.\,\ref{ml} the positions of the binary components are demonstrated in the $\log$\,$M$\,$-$\,$\log$\,$L$ diagram as well. This diagram was taken from \citeauthor{2014PASA...31...24E} (\citeyear{2014PASA...31...24E}). As can be seen from the figure, both components were found to be in agreement with the general $\log$\,$M$\,$-$\,$\log$\,$L$ correlation.

The distance of the system was calculated using the derived fundamental parameters, the interstellar absorption value $A_{v}$ (\citeauthor{2011ApJ...737..103S}, \citeyear{2011ApJ...737..103S}), and the bolometric correction (\citeauthor{2020MNRAS.496.3887E}, \citeyear{2020MNRAS.496.3887E}). As a result, the distance of the system was found to be 164\,$\pm$\,10 pc which is in agreement with the Gaia eDR3\footnote{https://gea.esac.esa.int/archive/} distance (156 pc).

As a result of the Fourier analysis of TESS light curve, we showed that the system does not exhibit any pulsational variation. Hence, we ruled out the pulsational characteristic of the primary component. However, when its position in the H-R diagram was examined, we found that the primary component of the system is placed inside the $\delta$\,Scuti instability strip (see, Fig.\,\ref{hr}). This is quite interesting because it is known that a significant amount of the stars pulsate inside this instability strip however some do not. \citeauthor{2019MNRAS.485.2380M} (\citeyear{2019MNRAS.485.2380M}) show that pulsator fraction reaches over 70\% in the middle of the $\delta$ Scuti instability strip. This number decreases through the cool and hot borders of the instability strip and the primary component of V948\,Her is placed very close to the cool border. It is now still a mystery why some stars inside the instability strip do not oscillate. To explain the existence of non-pulsators, there are some hypotheses such as wrong atmospheric parameters and the chemical peculiarity. However, for V948\,Her we have accurate atmospheric parameters and the chemical abundance distribution for the non-pulsating primary component. Hence, we simply rule out these hypotheses. During the research of possible pulsation frequencies, we reached micro magnitude level and still no pulsation frequency was found. One reason for the non-pulsator in the cool border of the $\delta$\,Scuti instability strip could be the convection. The primary component of V948\,Her is located nearly on the cool border where convection is very effective. Therefore, to accurately model the cool border of the $\delta$\,Scuti instability strip it would be better to understand the nature of pulsating and non-pulsating stars placed close to this area. For this reason, more samples of non-pulsating stars, with precise fundamental parameters, are requested. Therefore, eclipsing binary systems are unique because they allow us to determine sensitive fundamental parameters and examined their evolutionary status. V948\,Her is one of these systems that allow us to understand the reason why some systems inside the $\delta$ Scuti instability strip do not pulsate.





\begin{acknowledgements}
We would like to thank the reviewer for useful
comments and suggestions that helped to improve the
publication. This  study  has  been  supported by  the  Scientific  and  Technological  Research  Council (TUBITAK) project through 120F330. The authors thank Dr. Gerald Handler for discussions and his valuable suggestions.
The TESS data presented in this paper were obtained from the Mikulski Archive for Space Telescopes (MAST). Funding for the TESS mission is provided by the NASA Explorer Program. This paper makes use of data from the first public release of the WASP data \citep{2010A&A...520L..10B} as provided 
by the WASP consortium and services at the NASA Exoplanet Archive, which is operated by the California Institute 
of Technology, under contract with the National Aeronautics and Space Administration under the Exoplanet Exploration Program.
This research has made use of the SIMBAD data base, operated at CDS, Strasbourq, France.  This work has made use of data from the European Space Agency (ESA) mission \emph{Gaia} (https://www.cosmos.esa.int/gaia), processed by the \emph{Gaia} Data Processing and Analysis Consortium (DPAC, https://www.cosmos.esa.int/web/gaia/dpac/consortium). Funding for the DPAC has been provided by national institutions, in particular the institutions participating in the \emph{Gaia} Multilateral Agreement.
\end{acknowledgements}



\appendix                  

\begin{table}
\begin{center}
\caption[]{The minima times derived from the TESS, SuperWASP data and taken from O-C Gateway. I and II represent the primary and secondary minimums.}\label{minima_table}
 \begin{tabular}{lcc}
  \hline\noalign{\smallskip}
  Time of minima   & Minima type & Reference \\
  HJD (2400000+)  &              & \\
  \hline\noalign{\smallskip}
48501.1070   & I &  \citeauthor{1997ESASP1200.....E} (\citeyear{1997ESASP1200.....E}) \\
51318.0650   & I & 1  \\
53173.4690 $\pm$ 0.0004 & I & \citeauthor{2005IBVS.5588....1A} (\citeyear{2005IBVS.5588....1A})  \\
54943.4551 $\pm$ 0.0005 & I & \citeauthor{2009OEJV..107....1B} (\citeyear{2009OEJV..107....1B}) \\
55360.4494 $\pm$ 0.0005 & I & \citeauthor{2010IBVS.5943....1L} (\citeyear{2010IBVS.5943....1L})  \\
55365.5491 $\pm$ 0.0002 & I & \citeauthor{2010IBVS.5943....1L} (\citeyear{2010IBVS.5943....1L})\cite{2010IBVS.5943....1L}  \\
55376.3885 $\pm$ 0.0013 & II & \citeauthor{2010IBVS.5958....1L} (\citeyear{2010IBVS.5958....1L}) \\
56815.4574 $\pm$ 0.0005 & I & \citeauthor{2015OEJV..168....1H} (\citeyear{2015OEJV..168....1H})  \\
57139.3638 $\pm$ 0.0014 & I  & \citeauthor{2017OEJV..179....1J} (\citeyear{2017OEJV..179....1J}) \\
54642.50739 $\pm$ 0.00009 & I & SuperWASP  \\
54647.60967 $\pm$ 0.00030 & I & SuperWASP  \\
54656.53481 $\pm$ 0.00020 & I & SuperWASP  \\
54665.46272 $\pm$ 0.00020 & I & SuperWASP  \\
58984.58299 $\pm$ 0.00003 & I & TESS  \\
58997.33502 $\pm$ 0.00002 & I & TESS  \\
59008.81193 $\pm$ 0.00002 & I & TESS  \\
59011.36230 $\pm$ 0.00002 & I & TESS  \\
59021.56400 $\pm$ 0.00002 & I & TESS  \\
59034.31609 $\pm$ 0.00002 & I & TESS  \\
58983.94627 $\pm$ 0.00022 & II  & TESS \\
58994.14778 $\pm$ 0.00011 & II & TESS  \\
59008.17487 $\pm$ 0.00020 & II & TESS  \\
59010.72816 $\pm$ 0.00090 & II & TESS  \\
59023.47770 $\pm$ 0.00013 & II & TESS  \\
59033.67918 $\pm$ 0.00012 & II & TESS  \\
59405.40676 $\pm$ 0.00020 & I & this study  \\

  \noalign{\smallskip}\hline
\end{tabular}
\begin{itemize}
    \item 1. http://www.paschke.ch/Astronomie.htm
\end{itemize}
\end{center}  
\end{table}

\end{document}